\documentclass[12pt]{iopart}
\usepackage{graphicx}
\usepackage{color}
\usepackage{ulem}
\usepackage[colorlinks]{hyperref}

\begin{document}

\title[Bogoliubov excitation spectrum of an elongated condensate]{Bogoliubov excitation spectrum of an elongated condensate throughout a transition from quasi-one-dimensional to three-dimensional}
\author{Tao Yang$^{1,2}$, Andrew J Henning$^{1,3}$ and Keith A Benedict$^1$}
\address{$^1$School of Physics and Astronomy, University of Nottingham, Nottingham NG7 2RD, United Kingdom}
\address{$^2$Institute of Modern Physics, Northwest University, Xi'an 710069, P. R. China}
\address{$^3$Present Address: National Physical Laboratory, Teddington, Middlesex, TW11 0LW, United Kingdom}
\ead{yangt@nwu.edu.cn}

\begin{abstract}
The quasiparticle excitation spectra of a Bose gas trapped in a highly anisotropic trap is studied with respect to varying total number of particles by numerically solving the effective one-dimensional (1D) Gross-Pitaevskii (GP) equation proposed recently by Mateo \textit{et al.}. We obtain the static properties and Bogoliubov spectra of the system in the high energy domain. This method is computationally efficient and highly accurate for a condensate system undergoing a 1D to three-dimensional (3D) cigar-shaped transition, as is shown through a comparison our results with both those calculated by the 3D-GP equation and analytical results obtained in limiting cases. We identify the applicable parameter space for the effective 1D-GP equation and find that this equation fails to describe a system with large number of atoms. We also identify that the description of the transition from 1D Bose-Einstein condensate (BEC) to 3D cigar-shaped BEC using this equation is not smooth, which highlights the fact that for a finite value of $a_\perp/a_s$ the junction between the 1D and 3D crossover is not perfect.
\end{abstract}

\bigskip

\maketitle

\section{Introduction}
The character of elementary excitations is very important in understanding the macroscopic quantum behavior of a trapped Bose-condensed gas. Measurements of the collective modes in trapped gases of alkali-metal atoms\cite{PRL.77.420} were carried out soon after the discovery of Bose-Einstein condensates (BECs), and were followed by various theoretical investigations and numerical analyses\cite{PRL.77.2360,PRL.77.1671,dalfovo4,you,vic,PRA.56.5179,PRA.57.4669,PRA.66.043610,PRA.68.043610} with respect to different traps, especially regarding the low-energy excitations which are essentially of collective character. The tunability of magnetic and optical traps opens an extraordinary opportunity to study in practice not only one- and two-dimensional (1D, 2D) Bose systems, but also dimensional crossovers influenced by the number of particles, size and shape of the system, interaction strength, and temperature \cite{pos,PRL.105.265302,PRA.83.021605}. In particular in a very elongated harmonic trapping potential the condensate undergos a transition from quasi-1D BEC to three-dimensional (3D) cigar-shaped BEC with an increase of the number of atoms, or change in the trap aspect ratio (anisotropy) $\lambda$, where $\lambda=\omega_z/\omega_\perp$ with $\omega_z$ and $\omega_\perp$ being the axial and the radial frequencies of the trap potential respectively. Rapid progress in experimental techniques have made it possible to increase the aspect ratio of the trap, up to 1/2500 \cite{PRL.106.230405}, which makes many configurations possible. For many theoretical works and numerical simulations a general and accessible methodology for determining the static solutions and excitation frequencies of trapped BECs in different regimes is required.

The Gross-Pitaevskii (GP) mean-field theory\cite{gp1,gp2} has proven to be an indispensable tool in both analyzing and predicting the outcome of experiments of dilute condensates in the zero temperature limit. For a 3D system the GP equation is
\begin{equation} \label{GPE}
i\hbar\frac{\partial\psi(\textbf{r},t)}{\partial t}=-\frac{\hbar^2}{2m}\frac{\partial^2\psi(\textbf{r}, t)}{\partial\textbf{r}^2}+\big[V_{trap}(\textbf{r})+gN|\psi|^{2}\big]\psi(\textbf{r}, t),
\end{equation}
where $g=4\pi\hbar^2a_s/m$ is the 3D coupling constant with $a_s$ being the bulk $s$-wave scattering length and $m$ the mass of the atoms. The 3D order parameter $\psi(\textbf{r},t)$ for a gas of $N$ atoms is normalized to unity.
If a highly anisotropic harmonic trap potential, separable with respect to the $x, y$ and $z$ axes,
\begin{equation}
V_{trap}(\textbf{r})=V_\perp(x,y)+V(z)= \frac{1}{2}m\omega_\perp^2x^2+\frac{1}{2}m\omega_\perp^2y^2
+\frac{1}{2}m\omega_z^2z^2,
\end{equation}
is applied, the motion of atoms along the strongly confined directions is suppressed and thus the low energy degrees of freedom of the condensate in these dimensions is reduced. The condensate is in the ground state corresponding to motion in these directions, since they do not have enough energy to reach the related excited states. If the axial frequency of the trap is much smaller than the radial frequency, i.e. $\omega_z\ll\omega_\perp$, the corresponding dynamics of the systems become effectively 1D. The local density of the axial condensate, $\rho_{1D}$, takes a one dimensional form by integrating over the transverse coordinates,
\begin{equation}
\rho_{1D}(z, t)\doteq N\int d^2\textbf{r}_\perp|\psi(\textbf{r}, t)|^2,
\end{equation}
and the condensate wave function can be factorized in the form \begin{equation}
\psi(\textbf{r}, t)=\phi_\perp(\textbf{r}_\perp, \rho_{1D}(z, t))\phi(z, t),
\end{equation}
and $\rho_{1D}=N|\phi(z,t)|^2$ is obtained, providing the radial wave function $\phi_\perp$ is normalized to unity.

In the 1D limiting case the condensate dynamics are governed by the low dimensional GP equation where the coupling constant is $g_{1D}=g/2\pi{a_\perp^2}=2\hbar\omega_\perp a_s$ with $a_\perp=\sqrt{\hbar/m\omega_\perp}$. 
This 1D form of the GP equation is only applicable when the condition $a_s\rho_{1D}\ll 1$ is satisfied, where the mean-field
interaction energy can be treated as a weak perturbation. The condensate wave function minimizing the energy functional is, to the lowest order, the Gaussian ground state of the harmonic oscillator, and the condensate is tightly confined in the radial direction. This requirement is equivalent to a limit on the condensate occupation number to $N < a_\perp/a_s$ when $\hbar\omega_\perp$ exceeds the magnitude of the mean-field interaction energy.

The Gaussian approximation for the radial density profile becomes poor, however, when the number of particles or the aspect ratio of the trap increases, with the condensate becoming quasi-1D or 3D cigar-shaped. To find an efficient way to describe the axial dynamics of this kind of condensate with reduced dimensionality several authors have followed different routes to introduce more accurate analytical approximations which relate the radial and axial wave function with the 1D density profile\cite{sal,pet,ger,PRA.65.043614,jac,mat1,mat2,mun}. An effective-GP equation proposed recently by Mateo \textit{et al.} \cite{mun} is the most accurate and simple for condensates with repulsive interactions. For the attractive case one can refer Ref.\cite{PRA.65.043614}, which has a more complex interaction term. By deforming the interatomic interaction one can get an effective 1D-GP equation\cite{mat2,mun}
\begin{equation} \label{GPE-eff}
i\hbar\partial_t\phi(z,t)=-\frac{\hbar^2}{2m}\partial_z^2\phi(z, t)+\big[V(z)+\hbar\omega_\perp\sqrt{1+4a_sN|\phi|^2}~\big]\phi(z, t)
\end{equation}
which describes the two limiting cases and the crossover between them very well. The basic idea in this approach is the use of a local chemical potential $\mu_\perp=\hbar\omega_\perp\sqrt{1+a_s\rho_{1D}}$
to include the contribution of the radial degrees of freedom to the axial dynamics. This provides us a very convenient tool to investigate the excitations of elongated BECs in different regimes.

In this work we give a computationally  efficient way to calculate the Bogoliubov excitation spectra of trapped BECs ranging from 1D to 3D cigar-shaped configurations, including the crossover regime. By solving the Bogoliubov-de Gennes equations numerically, based on the effective-GP equation (equation (\ref{GPE-eff})), we get the static properties and quasiparticle excitation spectra over a range of atom numbers in a wide energy domain. We identify the range of parameters for which the effective 1D-GP equation is valid and find that the transition from the 1D BEC to the 3D cigar-shaped BEC is not smooth which highlights the fact that for a finite value of $a_\perp/a_s$ the junction between the two domains is not perfect \cite{PRA.66.043610,PRA.68.043610}. In the low energy excitation domain the simulation results are compared with the analytical results. In Sec.\ref{sec1} the ground state properties of the axial condensate obtained by numerically solving the effective 1D-GP equation (equation (\ref{GPE-eff})) are investigated. It is shown that the results are very accurate through comparisons with those calculated using the 3D-GP equation. In Sec.\ref{sec2} we present the Bogoliubov equations derived from the effective 1D-GP equation and the numerical methods we used. Analytical results for the very low lying excitations based on the variational method, hydrodynamical methods and the sum rule approach are discussed in Sec.\ref{sec3}. In Sec.\ref{sec4} the numerical results for the Bogoliubov spectra are compared with the analytical results given in Sec.\ref{sec3}.

In the numerical investigations we use $a_z=\sqrt{\hbar/m\omega_z}$ as the unit of length and $1/\omega_{z}$ as the unit of time. Hence we define dimensionless wavevectors $\tilde{k}=ka_{z}$, times $\tilde{t} = \omega_{\perp}t$ to minimize numerical instabilities and simplify our expressions.

\section{Ground state properties}\label{sec1}

In the Thomas-Fermi (TF) approximation the chemical potential obtained from the time-independent effective 1D-GP equation is
\begin{equation}
\tilde{\mu}=\frac{1}{\lambda}\left(1+\frac{1}{2}\lambda\tilde{Z}^2\right),
\end{equation}
where $\tilde{Z}$ is the dimensionless TF radius of the axial condensate, defined by the value of $\tilde{z}$ at which the equilibrium density $\rho_{1D}(\tilde z)$ vanishes. The density profile is then,
\begin{eqnarray}\label{den-eff}
\rho_{1D}(\tilde z)&=\frac{1}{4a_s}\left\{\left[\lambda\tilde{\mu}-\frac{1}{2}
(\sqrt{\lambda}\tilde{z})^2\right]^2-1\right\}\nonumber\\
&=\frac{1}{4a_s}(\sqrt{\lambda}\tilde{Z})^2\left[1-\bigg(\frac{\tilde z}{\tilde Z}\bigg)^2\right]
+\frac{1}{16a_s}(\sqrt{\lambda}\tilde{Z})^4\left[1-\bigg(\frac{\tilde z}{\tilde Z}\bigg)^2\right]^2
\end{eqnarray}
with $\rho_{1D}=0$ for $|\tilde z|\geq \tilde Z$. Then from the normalization condition
\begin{equation}
N = \int_{-\tilde Z}^{~\tilde Z}\rho_{1D}(\tilde z)d\tilde z,
\end{equation}
we obtain the relation which determines the axial half length
\begin{equation}\label{ax-length}
\frac{1}{15}(\sqrt{\lambda}\tilde{Z})^5+ \frac{1}{3}(\sqrt{\lambda}\tilde{Z})^3=N\sqrt\lambda\frac{a_s}{a_z}
=N\lambda\frac{a_s}{a_\perp}\doteq\chi.
\end{equation}
For a given  $\lambda$ the only relevant parameter is $\chi$ as defined in the above equation. An approximate solution of this equation is given by\cite{mat2}
\begin{equation}\label{axial-len}
\tilde{Z}=\frac{1}{\sqrt\lambda}\left[\frac{1}{(15\chi)^{4/5}+1/3}
+\frac{1}{57\chi+345}+\frac{1}{(3\chi)^{4/3}}\right]^{-1/4}.
\end{equation}

In the limiting case where $\chi\gg 1$
\begin{equation}
\tilde Z\approx \frac{1}{\sqrt{\lambda}}(15\chi)^{1/5}.
\end{equation}
This is the same result as gained with the 3D-TF approximation for a stationary 3D-GP equation, meaning that we have entered the 3D-TF regime. If $\chi\ll 1$, we obtain
\begin{equation}
\tilde Z \approx \frac{1}{\sqrt{\lambda}}(3\chi)^{1/3},
\end{equation}
which is the same result as that obtained from the stationary 1D-GP equation
using the TF approximation. It means that the 1D mean-field regime has been entered. For the local density approximation above to be valid, $\tilde Z\gg \tilde a_z$ must also be satisfied, which implies that $\chi\gg\frac{1}{3}\lambda^{3/2}$. This gives the boundary between the region where the TF axial density profile is valid and the Gaussian density profile (ideal 1D gas), is valid. If the condition $N\lambda \ll (a_s/a_\perp)^2$ \cite{PRA.66.043610} is satisfied, the Tonks-Girardeau (TG) regime \cite{gir} can be reached, however we will never enter this regime in this work.

\begin{figure}[t]
  \centering
   \includegraphics[scale=0.35]{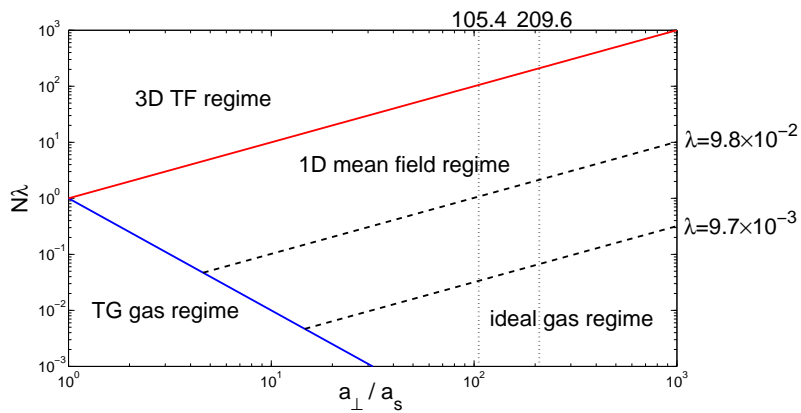}
   \caption{Phase diagram of the elongated condensate in the plane $N\lambda$ vs $a_\perp/a_s$. The dashed lines indicate the crossover region between the 1D mean-field regime and the ideal gas regime for $\lambda = 9.8\times 10^{-2}$ and $\lambda = 9.7\times10^{-3}$, respectively}
   \label{phase}
\end{figure}

Figure \ref{phase} shows a plot of $N\lambda$ vs $a_\perp/a_s$ along with several lines that schematically illustrate the regime that the gas is in. As the transition from one phase to another is gradual, these lines are only a guide as to where the crossover is. As $\chi$ decreases the gas will move from the 3D-TF regime into the 1D mean-field regime. The red solid line where $N\lambda=a_\perp/a_s$, i.e. $\chi = 1$, is marked to indicate the crossover region. As $\chi$ decreases further the gas will enter either the TG gas regime or the ideal gas regime. The ability to reach the ideal gas regime depends on the value of $\lambda$ and $a_\perp/a_s$. The blue solid line, $N\lambda=(a_s/a_\perp)^2$, shows the crossover between the TG gas regime, and either the 1D mean-field regime or the ideal gas regime. The ideal gas regime and the 1D mean-field regime are divided by the line $N\lambda = \frac{1}{3}\lambda^{3/2}a_\perp/a_s$. In figure \ref{phase} the location of this line is shown for two different values of $\lambda$ by the dashed lines.  We can see that the 3D-TF and the TG gas regions move further apart as $a_\perp/a_s$ becomes larger, leaving a larger parameter range for the 1D mean-field (the ideal gas) description. For certain values of $a_\perp/a_s$ and $\lambda$, the ideal gas regime becomes inaccessible. The system will not enter the TG regime for the parameters looked at in this work ($a_\perp/a_s = 105.4$, or $209.6$ when $N\lambda$ is greater than $10^{-3}$ as shown by the dotted lines in figure \ref{phase}).

In this work we consider an model in which $^{87}$Rb atoms are confined in a highly anisotropic harmonic trap. The trap frequency in the radial and axial direction is $\omega_\perp=2\pi\times 91$\textit{Hz} and $\omega_z=2\pi\times 8.9$\textit{Hz}, respectively. The 3D $s$-wave scattering length is $a_s=5.4$nm. The system then undergoes a transition from an 1D ideal gas, via a 1D mean-field gas and finally to a 3D cigar-shaped gas with increasing number of atoms.

\begin{figure}[t]
  \centering
   \includegraphics[scale=0.45]{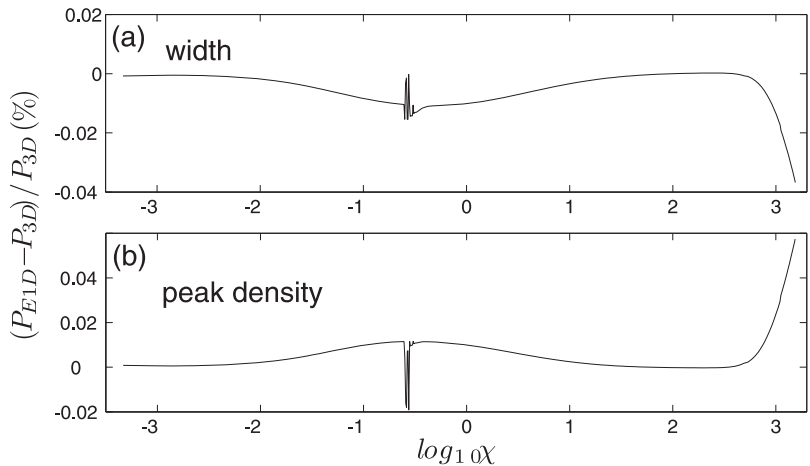}
   \caption{Relative differences of width (a) and peak density (b) of the axial condensate vs $log_{10}\chi$ calculated
   by the effective 1D- and 3D-GP equation.
   }
   \label{density-width}
\end{figure}

The ground state of the condensate is obtained via the imaginary-time-evolution grid method for the time dependent effective-GP equation (\ref{GPE-eff}) and the 3D-GP equation (\ref{GPE}). Theoretically the initial conditions we chose for the numerical simulation will only affect the efficiency of the calculation but not the final results. We found, however, that for the effective-GP equation (\ref{GPE-eff}), the choice of the initial chemical potential $\mu$ and the wave function during the transition between the 1D condensate and the 3D cigar-shaped condensate will greatly affect the accuracy of the ground state properties and then the excitation frequencies. We compare our simulation results obtained by using the effective 1D-GP equation (\ref{GPE-eff}) and the 3D-GP equation (\ref{GPE}). As shown in figure \ref{density-width}, the effective 1D-GP equation gives the peak density and the width of the condensate with accuracy better than 2$\%$ in all regimes up to some critical value of $\chi$. Especially in the two limiting cases the results are nearly the same as those of the 3D-GP equation. The most interesting finding is that the variation of these two quantities is not smooth from the 1D ideal gas to the 3D cigar-shaped condensate. In the crossover region there is a oscillation. We also find that there is a critical value of $\chi$ beyond which the deviation of the results increases very quickly, which means that the effective 1D-GP equation is not reliable anymore. This value is related to the aspect ratio of the trap. The larger the value of the aspect ratio the smaller the critical value. Beyond the critical value of $\chi$ the full 3D-GP equation must be applied to study the properties of the system.

\section{Bogoliubov-de Gennes equations}\label{sec2}

In the Bogoliubov approximation, the gas is initially assumed to be in thermal equilibrium and the Bose field
operator $\hat{\phi}(z, t)$ can be expanded linearly as
\begin{eqnarray} \label{Bog1}
\hat{\phi}(z, t)&=& [\phi_0(z) + \delta\hat\phi(z,t)]e^{-i\mu t/\hbar}\nonumber \\
&=&\phi_0\hat{a}_{0}+\sum_{j}[u_{j}(z)\hat{a}_{j}(t)
-v_{j}(z)^{*}\hat{a}_{j}^{\dag}(t)],
\end{eqnarray}
with $\phi_0$ being the time independent groundstate amplitude of the condensate, $\hat{a}_j^\dag, \hat{a}_j $ being the quasiparticle creation and annihilation operators, and where $\langle\hat{a}_{0}^{\dag}\hat{a}_{0}\rangle=N$ is the number of atoms in the condensate. $u_{j}$ and $v_{j}$ are the quasiparticle amplitudes. Substituting this expression into the effective 1D-GP equation, we can get the quasiparticle dynamical equation
\begin{equation}
i\hbar\partial_{t}\delta\hat\phi(z,t)=-\frac{\hbar^2}{2m} \partial_z^2\delta\hat\phi(z,t) +V_{eff}(z)\delta\hat\phi(z,t) + g_{eff}N\phi_0^2\delta\hat\phi^\dagger(z)
\end{equation}
where
\begin{equation}
V_{eff}(z) = \left\{ \begin{array} {cc} V(z)-\mu+2g_{eff}N|\phi_0(z)|^{2} &
 \textrm{1D} \\ \\ V(z)-\mu+\hbar\omega_\perp\sqrt{1+4a_sN|\phi_0|^2}+{g_{eff}N\phi_0^2} & \textrm{effective-1D} \end{array} \right.
\end{equation}
and
\begin{equation}
g_{eff}=\left\{ \begin{array} {cc} g_{1D} &
 \textrm{1D} \\ \\ \frac{g_{1D}}{\sqrt{1+4a_sN|\phi_0|^2}} & \textrm{effective-1D} . \end{array} \right.
\end{equation}
To get these results we have ignored all the higher order terms of $\mathcal{O}(\delta\hat\phi^2)$ and assumed that $\delta\hat\phi+\delta\hat\phi^\dagger$ is a very small quantity. The nonlinear term can be expanded as
\begin{equation}
\sqrt{1+4a_sN|\phi|^2}\approx\sqrt{1+4a_sN|\phi_0|^2}+ \frac{2a_sN\phi_0}{\sqrt{1+4a_sN|\phi_0|^2}} (\delta\hat\phi+\delta\hat\phi^\dagger).
\end{equation}
It is obvious that the 1D result is nothing but the limiting case of the effective-1D result.

It is then easy to show that $u_j(z)$ and $v_j(z)$ fulfill the Bogoliubov equations,
\begin{equation}\label{bog}
\left(
\begin{array}
[c]{cc}
A(z) & -B(z)\\
B^{\ast}(z) & -A(z)
\end{array}
\right) \left(
\begin{array}
[c]{c}%
u_{j}\left(  z\right)  \\
v_{j}\left(  z\right)
\end{array}
\right)  =E_j\left(
\begin{array}
[c]{c}%
u_{j}\left(  z\right)  \\
v_{j}\left(  z\right)
\end{array}
\right),
\end{equation}
where the characteristic matrix elements are
$A(z)=-\frac{\hbar^2}{2m}\partial_z^2+V_{eff}(z)$ and $B(z)=g_{eff}N\phi_0^{2}( z)$. Imposing the Bose commutation
rules on the operators $\hat{a}_j^\dag$ and $\hat{a}_j $, we find that the quasiparticle amplitudes $u$ and $v$ must
obey the orthogonality conditions
\begin{equation}\label{norm1}
\int dz [u_{j}^{*}u_{k}-v_{j}^{*}v_{k}]=\delta_{jk},~~~     \int dz [u_{j}^{*}v_{k}-u_{k}^{*}v_{j}]=0.
\end{equation}
In general, we can expand the mode functions $u(z)$, $v(z)$ and the condensate wave function $\phi(z)$ on a basis set $\varphi_n(z)$,
\begin{equation}
u_j(z)=\sum_{n=0}^{\infty}a^{(j)}_{n}\varphi_{n},~
v_j(z)=\sum_{n=0}^{\infty}b^{(j)}_{n}\varphi_{n},~
\phi_j(z)=\sum_{n=0}^{\infty}c^{(j)}_{n}\varphi_{n},\nonumber
\end{equation}
Of course, in practice we can only involve a limited number of basis functions in our simulation.

For an effective-1D BEC described by equation (\ref{GPE-eff}), we denote the nonlinear interaction term as
\begin{equation}\label{int_expansion}
\tilde{\Lambda}_{n,n'}=\tilde{g}_{eff}\int
d\tilde{z}\sum_{l,l'}c^{(j)}_{l}c^{\ast(j)}_{l'}\tilde \varphi_{n}(\tilde{z})\tilde \varphi_{l}(
\tilde{z})\tilde \varphi_{l'}(\tilde{z})\tilde \varphi_{n'}(\tilde{z}),
\end{equation}
and then equation (\ref{bog}) can be rewritten as
\begin{equation}\label{bog7}
 \left(\begin{array}[c]{cc}%
{X}_{n,n'} & {-Y}_{n,n'}\\
{Y}_{n,n'}^{\ast} & {-X}_{n,n'}%
\end{array}
\right)\left(\begin{array}{c} a^{(j)}_{n'} \\ b^{(j)}_{n'}
\end{array}\right)=\tilde E_{j} \left(\begin{array}{c} a^{(j)}_{n'} \\ b^{(j)}_{n'}
\end{array}\right),
\end{equation}
where $\tilde{g}_{eff}=g_{eff}/(a_z\hbar\omega_z)$, $\tilde E_j= E_j/\hbar\omega_z$, $X_{n,n'}=\left(\varepsilon_n-\tilde\mu\right)\delta_{n,n'}+2\tilde{\Lambda}_{n,n'}$, $Y_{n,n'}=\tilde{\Lambda}_{n,n'}$, and $\tilde E_j$ is the eigenvalue of Bogoliubov equations. The value of $\varepsilon_n$ depends on the specific form of the basis function $\varphi_n$. If we choose the simple-harmonic-oscillator (SHO) basis, which are the eigenstates of $\tilde{H}_{0}=-\partial_{\tilde{z}}^2/2+\tilde{V}(\tilde{z})$, then we get $\tilde{H}_0\tilde \varphi_n=(n+1/2)\tilde \varphi_n=\varepsilon_n\tilde \varphi_n$ with $n=0,1,\ldots$ being the corresponding quantum number. Another choice is the plane-wave (PW) basis, which is computationally the simplest (by virtue of the Fast Fourier transform method) in which $\varepsilon_n = \tilde{k}_n^2/2+\tilde{V}^{k_n}$ with $\tilde{k}_n$ being the dimensionless momentum and $\tilde{V}^{k_n}$ the Fourier part of $\tilde{V}=\tilde{z}^2/2$.


From equation (\ref{bog7}), we can get the condition for a nonzero solution is
\begin{equation}\label{Bog2}
\left|X_{n,n'}^{2}-Y_{n,n'}^{2}-\frac{1}{\lambda}\tilde E_j^{2} I\right| =0
\end{equation}
with $I$ being a unit matrix.

The quantum number $n$ is defined as the number of nodes found in the axial wave function $\varphi_n(z)$. If there exists a solution of the normal-mode equations for positive $\tilde{E}_j$, then a negative solution also exists because if we replace $\tilde{E}_j$ by $-\tilde{E}_j$ and $u_j$ by $v_j$, the Bogoliubov equations (\ref{bog}) remain unchanged.

A formal solution of (\ref{bog}) at zero energy is $\tilde{E}_0=0$ with $u_0(z)$ and $v_0(z)$ being proportional to the ground-state wave function of the condensate, but the norm of this solution is zero which is not normalizable in this way (\ref{norm1}). The existence of a zero mode is consistent with Goldstone's theorem \cite{gold} since the nonzero average of the wavefunction breaks the U(1) global gauge symmetry of the Hamiltonian.

To solve equations (\ref{bog}) or (\ref{bog7}) we need to get the ground state wave function of the condensate by solving the effective 1D-GPE numerically. Then the
coefficients $a_n$, $b_n$ and the excitation energy, $\tilde{E}_j$, are calculated by matrix diagonalization with a given basis set.

The accuracy of the SHO method depends strongly on the number of basis states retained. The higher the energy modes we want to get, the more basis states we need to use. However, the largest number of basis states we can use are limited by the computational cost of solving the Hermite polynomials. So the number of excitation modes that can be calculated is limited. Another limitation of this method is that to get accurate values for $\tilde{\Lambda}$, the integrand in equation (\ref{int_expansion}) must be expressed as a polynomial. The basis functions $\varphi_n(z)$, for a harmonic trapping potential $\tilde V(\tilde z)$, can be given by the dimensionless SHO functions
\begin{equation}
\tilde\varphi_n(\tilde z)=\left[\frac{1}{\sqrt{\pi}2^nn!}\right]^{1/2} e^{-{\tilde z^2}/{2}}H_n(\tilde z).
\end{equation}
By introducing another variable $\zeta=\tilde z/\sqrt{2}$ and redefining the Hermite polynomial, $H_n(\tilde z)$, the effective 1D interacting term $\Lambda_{n,n'}$ turns out to be an integral
\begin{equation}\label{g-int}
\tilde\Lambda_{n,n'}=\sum_{l,l'} \frac{c^{(j)}_{l}c^{\ast(j)}_{l'}}{\sqrt{2}}\int
d\zeta \frac{1}{\lambda}\tilde{g}_{eff}e^{-\zeta^2}H_{n}(\zeta)H_{l}(
\zeta)H_{l'}(\zeta)H_{n'}(\zeta).
\end{equation}
which can be evaluated by using a gaussian quadrature integration technique\cite{wil}. For the 1D-GP equation, $\tilde{g}_{eff}$ is just a constant which can be taken out of the integral, while for the effective 1D-GP equation it is a function of the ground state wave function. Hence this method is not very accurate for the effective 1D-GP equation (\ref{GPE-eff}).

The PW expansion, related to the coordinate space, requires many more modes than the SHO basis. The size of the system and the mesh spacing are set when we calculate the ground-state wave function. The number of excitation modes we get is the same as the number of mesh points, but only the first $20\%$ are stable (accurate) when we change the size or number of mesh points. However, we can still get very high energy excitation modes because of the large number of grides we used. In the following discussions we will use the numerical results corresponding to a PW basis.

\section{Theoretical Analysis of collective excitations}\label{sec3}

In a uniform 3D gas, the amplitudes $u$ and $v$ are plane waves and the resulting dispersion law for the elementary excitations takes the famous Bogoliubov form \cite{bec-book}
\begin{equation}
E(p)=\hbar\omega(p)=\left[\frac{g\rho}{m}p^2+\left(\frac{p^2}{2m}\right)^2\right]^{1/2},
\end{equation}
where $p$ is the momentum of the excitation and $\rho$ is the density of the gas. For small momenta the spectrum takes the phonon-like form $E(p)=cp$, with $c=\sqrt{g\rho/m}$ being the sound velocity. At large momenta the quasiparticle behaves like free particle with energy $E(p)\approx p^2/2m+g\rho$.

\subsection{Variational analysis of the low-lying excitations in the low density domain}

For a 1D-BEC the spectrum of the low energy excitations can be studied by following a time-dependent variational method introduced in Ref.\cite{vic} for the 3D case.

Let's start with a dimensionless time-dependent GP equation
\begin{equation}\label{v1}
i\partial_{\tilde{t}}\tilde\phi(\tilde{z}, \tilde{t})= \left[
-\frac{1}{2}\partial_{\tilde{z}}^2+\tilde{V}(\tilde{z})
+\tilde{g}_{1D}N|\tilde\phi|^2\right]\tilde\phi(\tilde{z},\tilde{t}).
\end{equation}
The problem of solving equation(\ref{v1}) can be restated as a variational problem corresponding to the minimization of the action related to the Lagrangian density
\begin{equation}\label{v2}
\mathcal{L}=\frac{i}{2}\big(\phi\partial_{\tilde{t}}\tilde{\phi}^*
-\tilde{\phi}^*\partial_{\tilde{t}}\tilde{\phi}\big)
+\frac{1}{2}\partial_{\tilde{z}}\tilde{\phi}^*\partial_{\tilde{z}}\tilde{\phi}
+\tilde{V}(\tilde z)|\tilde{\phi}|^2+\frac{\tilde{g}_{1D}}{2}N|\tilde{\phi|}^4.
\end{equation}

It is natural to choose a gaussian ansatz wave function
\begin{equation}\label{v3}
\tilde\phi(\tilde{z},\tilde{t})=A(\tilde{t})
e^{-\frac{[\tilde{z}-\alpha(\tilde{t})]^2}{2\sigma(\tilde{t})^2}
+i\tilde{z}\beta(\tilde{t})+i\tilde{z}^2\gamma(\tilde{t})}
\end{equation}
in the very weak interaction regime, and the normalization is given by $\int|\tilde{\phi}|^2d\tilde{z}=1$. Here $\alpha, \beta, \gamma$ and $\sigma$ are all real parameters. The imaginary terms of the exponential are introduced to make the results reliable \cite{vic}. We insert equation (\ref{v3}) into equation (\ref{v2}) and calculate an effective Lagrangian, $L$, by integrating the Lagrangian density over the whole coordinate
space $L=\int_{-\infty}^{\infty}\mathcal{L}d\tilde{z}$. Through a long but straightforward calculation, we get the equations of motion
\begin{eqnarray}
\partial_{\tilde{t}}^2{\alpha}+\alpha &=0 \label{v4}\\
\partial_{\tilde{t}}^2{\sigma}+\sigma
&=\frac{1}{\sigma^3}+\frac{G}{\sigma^2}\label{v5}.
\end{eqnarray}
The fixed points of equations (\ref{v4})-(\ref{v5}) can be obtained from
\begin{eqnarray}
\alpha_0 = 0, \\
\sigma_0-\frac{1}{\sigma_0^3}-\frac{G}{\sigma_0^2}=0
\end{eqnarray}
where $G=\tilde{g}_{1D}N/\sqrt{2\pi}$. There are two solutions for $\sigma$, one positive and one negative. As $\sigma$ controls the width of the Gaussian (see equation (\ref{v3})) only the positive one is interesting. Expanding equations (\ref{v4}) and (\ref{v5}) around this equilibrium point, we can obtain the following frequencies for low-energy excitation modes:
\begin{eqnarray}
E_1 &= \hbar\omega_z,\label{v6}\\
E_2&=\left[1+2(\frac{3}{2\sigma^4}
+\frac{G}{\sigma^3})\right]^{1/2}\hbar\omega_z\label{v7},
\end{eqnarray}
where we have restored all the units.

$E_1$ corresponds to the dipole oscillation characterizing the motion of the centre of mass unaffected by the interatomic force. The oscillation frequency is just the frequency of
the harmonic trap in the axial direction due to the
harmonic confinement, which indicates that the motion of the centre of mass can be exactly decoupled from the internal degrees of freedom of the system.

$E_2$, the axial breathing mode, is the excitation energy corresponding to the width and the interaction strength of the condensate. The width of the condensate increases with the strength of the interaction between the particles. We note that equation (\ref{v7}) is only applicable in a restricted region where an ideal gas condensate with a Gaussian density profile is realized. In this region, $E_2$ decreases with increasing $G$ because the condensate expands rapidly in the axial direction.

\subsection{Analytical solutions in the hydrodynamic limit}

The analytic solutions of the linearized GP equation in the hydrodynamic limit for the low-energy axial modes were obtained in Refs.\cite{str1,lin}. The excitation frequency of the $j_{th}$ mode is given by
\begin{eqnarray}
\omega_{j-1D} &= \sqrt{j(j+1)/2}~\omega_z,\label{1d-exc}\\ \omega_{j-3DTF} &= \sqrt{j(j+3)/4}~\omega_z, \label{3d-exc}
\end{eqnarray}
for the 1D mean-field and 3D-TF regimes respectively, where
$j$ is a positive integer indicating the different excitation modes. These two equations are only valid if $\omega_j\ll\omega_\perp$. For noninteracting particles in a harmonic potential the 1D form of the frequencies of the excitations are given by $\omega = j\omega_z$. The dispersion relation of the normal modes of the condensate is changed significantly from the noninteracting behaviour, as a consequence of two-body interactions. Notice that in the case $j=1$, the excitation frequency coincides with the oscillator frequency in the two limiting cases. In accordance with the general considerations discussed in the previous section, this also holds for the crossover regime. The results of these low-energy modes, which can be directly excited by suitable modulation of the harmonic trap, have been confirmed to high precision by experiments\cite{mew,andrews}. The fact that the frequency of the first excitation is not affected by the interatomic interactions, demonstrating that the lowest excitation in the system is the dipolar oscillation, offers a direct test on the numerical accuracy of our calculations.

\subsection{Collective excitations in the high density domain}

Following a sum-rule approach \cite{PRL.77.2360,PRL.77.1671,dalfovo,you}, the collective frequency in the high density domain (from the 1D mean-field regime to the 3D-TF regime) can be derived by using the formula \cite{PRA.66.043610}
\begin{equation}\label{v8}
\omega^2=-2\frac{\langle z^2\rangle}{d\langle z^2\rangle/d\omega_z^2},
\end{equation}
where $\langle z^2\rangle\doteq N^{-1}\int_{-Z}^{Z} dzz^2|\phi_0(z)|^2$ is the expectation value of the square radius. This equation gives the same bounds to the frequency of the axial breathing mode as equations (\ref{1d-exc}) and (\ref{3d-exc}). With equations (\ref{den-eff}) and (\ref{axial-len}), we obtain
\begin{equation}\label{exc2}
\frac{\omega^2}{\omega_z^2}=\frac{4(\sqrt\lambda\tilde Z)^3-15\chi[(\sqrt\lambda\tilde Z)^2+5]}{3(\sqrt\lambda\tilde Z)^3-6\chi[(\sqrt\lambda\tilde Z)^2+5]},
\end{equation}
which means that the excitation frequency can be related solely to the axial half-length of the elongated condensate in the equilibrium configuration because $\chi$ is also a function of $\tilde Z$ (see equation (\ref{axial-len})). This provides a straightforward method to measure the excitation frequency experimentally.

\begin{figure}[t]
  \centering
   \includegraphics[scale=0.45]{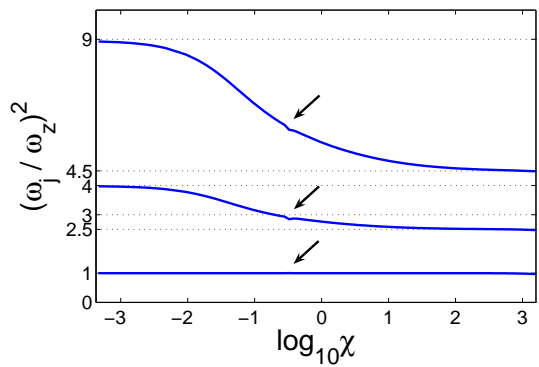}
    \caption{  The frequency of the three lowest excitation mode with respect to $log_{10}\chi$. (curves from bottom: $j = 1,~2,~3$.)  
    The arrows indicating the transition between the low density domain and the high density domain (see details in text). }\label{sg3} 
\end{figure}

\section{Numerical results and discussion}\label{sec4}

In figure \ref{sg3}, we show the Bogoliubov spectra of the three lowest excitations with respect to the parameter $\log_{10}{\chi}$. The first excitation frequency is always $\omega_z$ as predicted by the theory, while the frequencies of the second and the third excitation modes decrease gradually with $\chi$. The upper and lower bound of the axial breathing mode are $\sqrt{3}\omega_z$ and $\sqrt{2.5}\omega_z$ respectively, as calculated by equations (\ref{1d-exc}) and (\ref{3d-exc}) when $j=2$, while the frequency for the ideal gas limiting case is $2\omega_z$. The relative error of the frequency of the dipole mode is less than $0.15\%$, implying a high degree of accuracy in our simulations. Our simulation results also show a very good performance of recovering the same bounds of 1D and 3D cigar-shaped condensate limiting cases for the second and the third excitation modes. However, the bounds for the 1D mean-field condensate, $\omega_2=\sqrt{2.5}\omega_z$ and $\omega_3=\sqrt{6}\omega_z~...$, are not completely recovered. Instead, there is a kink between the two domains (see arrows in figure \ref{sg3}). We shall explain this later.

\begin{figure}[t]
  \centering \includegraphics[scale=0.45]{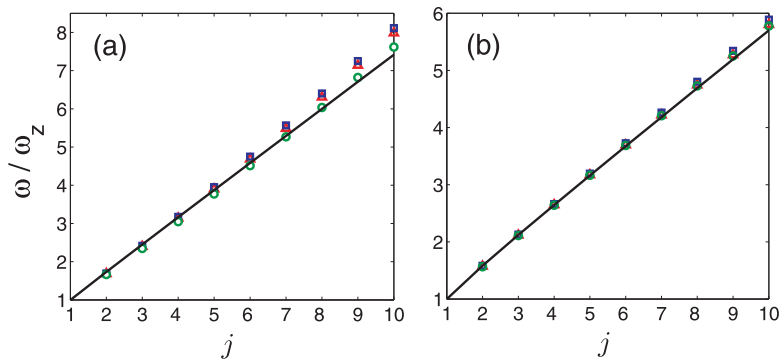}
  \caption{ (a) The energy spectrum for the 1D mean-field regime. The solid curve is the theoretical prediction from equation (\ref{1d-exc}) and open symbols are numerical results (open squares, triangles and circles correspond to $\chi = 0.47, ~0.51, ~0.93$, respectively). (b) The energy spectrum for the 3D-TF regime. The solid curve is the theoretical prediction from equation (\ref{3d-exc}) and the open symbols are the numerical results. (open squares, triangles and circles correspond to: $\chi = 4.43\times10^2, ~1.31\times10^3, ~2.33\times10^3$) 
  }\label{energy-compare}
\end{figure}

As shown in figure \ref{energy-compare}, for the 1D mean-field and the 3D-TF limiting cases, the frequencies of the very low-lying excitations with different interaction strengths are so close that they can be well approximated by a function of the excitation modes $j$ ( equations (\ref{1d-exc}) and (\ref{3d-exc}) ), as if they do not depend on the interatomic interaction. This differs from the uniform case where the dispersion relation,
in the corresponding phonon regime, depends explicitly on the interaction through the velocity of sound. For a fixed value of $\chi$ the accuracy of predictions (\ref{1d-exc}) and (\ref{3d-exc}) decreases as $j$ increases, and in the 1D mean-field regime, the frequency of the low-lying excitations is more sensitive to the variation of $\chi$ than in the 3D-TF regime.

\begin{figure}[t]
  \centering
  \includegraphics[scale=0.45]{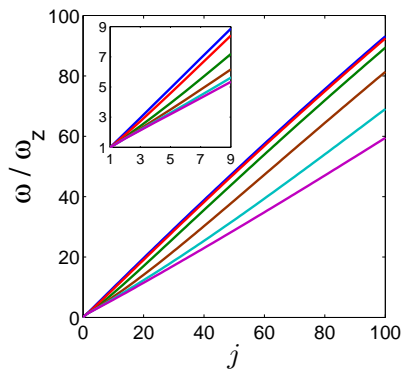}
  \caption{ The energy spectrum vs the excitation modes. $\omega_z=2\pi\times 8.9$Hz, $\omega_\perp=2\pi\times 91$Hz.
  (curves from top N=10, 100, $10^3$, $10^4$, $10^5$, $10^6$.) Inset is a zoom plot for the very low-lying excitations.}
  \label{spectra} 
\end{figure}

In figure \ref{spectra} the excitation spectra for different values of $\chi$ are plotted for one hundred excitation modes. The frequency of the excitations increases monotonically with $j$ for a given $\chi$, while for a given $j$ the frequency increases as $\chi$ is reduced. The different spectra diverge as $j$ increases. In the 1D mean-field regime, this divergence decelerates with increasing $j$, whereas for 3D-TF regime it accelerates.

\begin{figure}[]
  \centering
  \includegraphics[scale=0.45]{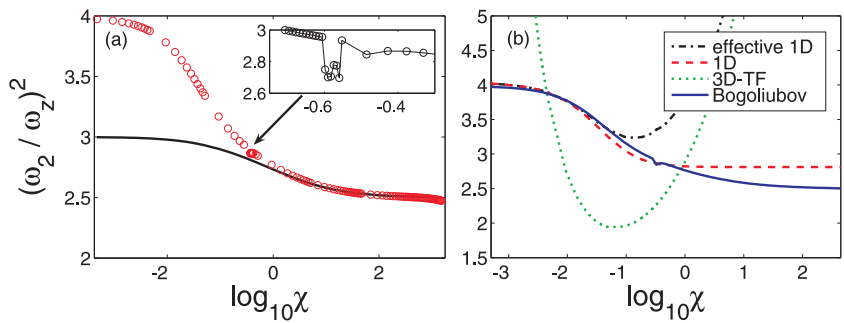}
  \caption{The frequency of the axial breathing mode. (a) The solid line is the theoretical prediction from equation(\ref{exc2}). Open circles are numerical results. (b) The dashed line, dash dotted line and dotted line are the theoretical predictions from equation(\ref{v7}) by using the width of the condensate which is obtained from the 1D-GP equation, effective 1D-GP equation and 3D-TF approximation, respectively. $\omega_z=2\pi\times 8.9$Hz, $\omega_\perp=2\pi\times 91$Hz. }\label{exc3}
\end{figure}

In the low density domain (from the ideal gas regime to the 1D mean-field regime) the theoretical prediction of the frequency of the axial breathing mode has been obtained as equation (\ref{v7}) by the variational analysis. In figure \ref{exc3} we use this equation and the width of the condensate in the axial direction calculated by the effective 1D-GP equation, 1D-GP equation and 3D-TF approximation to compare the results with the results given by the Bogoliubov calculation. Both the results from the effective 1D-GP equation and the 1D-GP equation work well in this region.

We use this equation and the axial half length of the condensate calculated through the effective 1D-GP equation to obtain the frequency of the excitations in this region. figure \ref{exc3} shows that our simulation is in good agreement with this prediction.

As shown in figure \ref{sg3}, the Bogoliubov spectra satisfy the same bounds for the ideal gas regime and the 3D-TF regime. However, there is a kink between the two domains (see arrows in figure \ref{sg3}). The zoom plot of this region is in figure \ref{exc3}(a). The limit value of the square mode frequency of the 1D mean-field regime is $\omega^2=3$ (see equation (\ref{1d-exc})). As noted in Refs.\cite{PRA.66.043610,PRA.68.043610} the 1D mean-field regime can only be identified provided $a_\perp/a_s\ll 1$. This means that if we decrease the density (decreasing $log_{10}\chi$) from the high density domain then we will reach this limit from the right hand side as shown in figures \ref{sg3} and \ref{exc3}. On the contrary, if we increase the density (increasing  $log_{10}\chi$) from the low density domain then we will reach the 1D mean-field limit from the left hand side. So there is a junction between the two domains originating from the finite value for $a_\perp/a_s$. Keeping in mind that $\lambda\ll1$, then in order to have a perfect junction at $(\omega_2/\omega_z)^2 = 3$, i.e. this value can be fully reached from both sides of the 1D mean-field regime, we should take the limit $a_\perp/a_s\rightarrow\infty$. For $a_\perp/a_s\simeq1$ one would have a smooth transition between these two domains, without any plateau. From the analysis in Sec. 4 we know that the excitation frequency can be solely related to the axial half-length (see equations (\ref{v7}), (\ref{v8}) and (\ref{exc2}) ) or the peak density \cite{PRA.68.043610} of the elongated condensate in the equilibrium configuration. So there should be bounds for the axial width and the peak density of the condensate in the transition area as well. As shown in figure \ref{density-width} our results from employing the effective 1D-GP equation revealed that the variation of peak density and width of the condensate is not a smooth curve but instead oscillates in the transition region. In Sec. 4 we note that equation (\ref{exc2}) is valid in the high density domain, while equation (\ref{v7}) is valid only in the low density domain. We compare the numerical results with the theoretical predictions from equations (\ref{exc2}) and (\ref{v7}) for the two domains in figure \ref{exc3} and find they coincide very well, except when the system approaches the 1D mean-field regime. The validity of the 1D-GP equation is strongly related to the value of $\chi$. In figure \ref{exc3}(a), we can see that the frequency of the excitation decreases dramatically when $\log_{10}\chi > 3$, meaning that this method is not applicable further into this region.

\section{Conclusions}\label{sec5}

In this paper we have calculated the Bogoliubov excitation spectra of a Bose gas in a very elongated trap. The system undergoes a 1D ideal gas, 1D mean-field gas and 3D cigar-shaped gas transition with increasing number of atoms or increasing aspect ratio of the trap. In order to get the excitation spectrum in all of these regimes, we take the effective 1D-GP equation developed by Mateo \textit{et al.} \cite{mat2,mun} as a starting point. The Bogoliubov equations are solved by using the matrix diagonalization method with a plane wave basis. The results of our simulations are compared with those from the variational analysis in the low density domain, and the sum rule approach in the high density domain. We find that the Bogoliubov method fails to give the accurate spectrum in the transition region between these two domains. The plateau area is replaced by a kink with an irregular trajectory. We also find that there is a critical value of $\chi$ where the effective 1D-GP equation is not applicable any more. In this work we only consider the mean-field approach in which thermal and quantum fluctuations are negligible. However, to compare with experimental results we have to consider the effects of finite temperature and quantum fluctuations on the system. Then one can use the truncated Wigner method \cite{twa1, twa2, twa3, JPB.46.145307} to solve the effective 1D-GP equation numerically. A recent experiment \cite{PRA.83.021605} studied the dimensional transition from 1D- to 3D-condensates when adjusting temperature of the system. The excitation properties of the above system would be interesting to pursue in further studies.

\section*{Acknowledgment}

We thank German Sinuco and Bo Xiong for many useful discussions. We also acknowledge support from the EPSRC. T.Y. acknowledges support through NSFC11247605 and NSFC11347025.

\section*{References}
\bibliographystyle{unsrt}
\bibliography{Bogoliubov_excitation_arxiv.bbl}

\end{document}